\documentclass[sigconf, nonacm]{acmart}

\usepackage{graphicx}
\usepackage{xspace}
\usepackage{url}
\usepackage{subcaption}

\newcommand{\CE}{\textsf{CardEst}\xspace}

\newcommand{\baihe}{\textsf{Baihe}\xspace}
\newcommand{\postgres}{\textsf{PostgreSQL}\xspace}

\mathchardef\mhyphen="2D

\definecolor{mygrey}{RGB}{230,230,240}

\begin{document}
	\title{Baihe: SysML Framework for AI-driven Databases}
\author{Andreas Pfadler$^1$, Rong Zhu$^1$, Wei Chen$^1$, Botong Huang$^1$, Tianjing Zeng$^{1, 2}$, \break Bolin Ding$^1$, Jingren Zhou$^1$}
	\vspace{0.5em}
	\affiliation{%
	\institution{\LARGE{\textit{$^1$Alibaba Group}, \textit{$^2$Renmin University of China}} \\
	\textsf{\{andreaswernerrober, red.zr, wickeychen.cw, botong.huang, zengtianjing.ztj, \break bolin.ding, jingren.zhou\}@alibaba-inc.com}
	}
	\vspace{0.5em}
	}

\begin{abstract}
We present \baihe, a SysML Framework for AI-driven Databases. Using \baihe, an existing relational database system may be retrofitted to use learned components for query optimization or other common tasks, such as e.g. learned structure for indexing. To ensure the practicality and real world applicability of \baihe, its high level architecture is based on the following requirements: separation from the core system, minimal third party dependencies, Robustness, stability and fault tolerance, as well as stability and configurability. Based on the high level architecture, we then describe a concrete implementation of \baihe for \postgres and present example use cases for learned query optimizers. To serve both practitioners, as well as researchers in the DB and AI4DB community \baihe for \postgres will be released under open source license.
\end{abstract}

\maketitle

\section{Introduction}
\label{sect:intro}
Learned systems, particularly learned database systems, have been a recent research hot spot~\cite{zhou2020database, han2021CEbenchmark, kraska2019sagedb}. As part of this trend, a considerable number of researchers have proposed integration of machine learning models into relational database systems in order to optimize overall system performance and cope with the ever increasing amounts of data that are being processed in modern data intensive applications.

As part of this trend, machine learning based methods for a wide range of problems have been proposed. Most efforts appear to have been spent on problems related to query planning and optimization, where machine learning models may be used e.g. for cardinality estimation~\cite{han2021CEbenchmark, wu2020bayescard, hilprecht2019deepdb, yang2019deep, zhu2020flat}, join order selection~\cite{marcus2018deep}, cost prediction~\cite{siddiqui2020cost, vu2021learned} or query planner steering through hint set proposals~\cite{marcus2020bao}. 

In addition to the above, other lines of research have focused on learned data structures~\cite{kraska2018case} used for efficiently querying large indices~\cite{nathan2020learning, marcus2018deep, perera2021no, sabek2021case}, automatic database configuration tuning, accurate query run time prediction and other problems~\cite{zhou2020database}.

As is the case with many publications in the machine learning space, on paper the reported results do indeed look impressive and suggest significant performance gains, if such methods were ever deployed in the real world. Yet, as most prior research has been justified on the basis of numerical experiments, performed \textit{outside} of real systems, there has been a striking absence of work focusing on the practical deployment of ML-based methods \textit{inside} real world systems. The central goal of this work is to present a concrete proposal on how this notable gap between theory and practice may be closed. For this aim, we propose here the \baihe SysML framework.

\textbf{\baihe} is a general framework developed for integrating machine learning models into a relational database systems, meaning that its fundamental architecture should provide a blueprint for other implementation projects in the AI4DB and SysML space. To exemplify its applicability and generate some more robust evidence for the actual usefulness of prior research in AI4DB, we have developed the \baihe extension for \postgres, a widely used, highly successful and extensible relational database system. 

This article is now structured as follows: In Section \ref{sect:background}, we first discuss general challenges for the implementation of AI-driven databases and then present our long term vision for AI4DB. Based on this, we then describe the rationale and fundamental principles for \baihe's design.

Later on, in Section \ref{sect:design} we present the concrete design and features of \baihe in the case of \postgres and further discuss implementation specifics. After show a typical example use case for \baihe on query optimizer in Section~\ref{sect:example}, we conclude this article in Section \ref{sect:conclusion}.

\section{Background and Design Rationale}
\label{sect:background}
On a fundamental level, the nature of and requirements on a database system appear to be at odds with the inherently stochastic nature of machine learning. Database developers, users, and administrators expect rock-solid, stable and most of all deterministic behavior. On the other hand, machine learning models may produce predictions with hard to estimate error bounds and their generalization ability may fail catastrophically, once the input data distribution changes. Moreover, failures may not be readily detectable, since they might only be visible through a change in some numerical value rather than a concrete and meaningful error message. 

Moreover, since training of a machine learning model generally amounts to solving a mathematical optimization problem, often through the use of stochastic optimizers, model training is difficult to automate, requires close supervision by experts and may thus not be fully regarded as a well defined process. Hyperparameters, useful termination criteria and evaluation guidelines all need be carefully developed on both a per-model and per-dataset basis. All of this makes it very difficult to integrate automated training procedures into a system with strong requirements on robustness and stability.

As challenging as this may sound, we still believe that the practical combination of machine learning and databases into the next generation of database systems is not a hopeless endeavor. 

To see this, we first note that some core components of a database system \textit{already} rely on \textit{statistical} estimates. Modern query optimizers in particular rely on such estimates e.g. in the context of cardinality and cost estimation, which are in turn used for join-order-selection and other automated decision making. The vision outlined by e.g. SageDB~\cite{kraska2019sagedb} has made a compelling case for using more advanced probabilistic methods - powered by machine learning - to improve existing issues arising from the relatively simple statistical estimations which are still part of the standard implementations of today's database systems.

Second, we believe that issues related to using machine learning enabled components inside a database system, should merely be regarded as \textit{additional requirements} on the engineering of such systems. Hence, we argue that one should in principle be able to satisfactorily fulfil these  requirements, \textit{if} they are properly taken into account during the system design phase and possibly enhanced through system-algorithm co-design. We hope to exemplify and support this point of view through the \baihe framework presented in this work.

\subsection{Learned Databases: Brief Overview}
\label{sect:learned_db}
Before presenting the high-level architecture of \baihe and the rationale behind it, we first review related literature and describe the current stage and a future vision for learned database systems. 

We categorize learned databases into four levels, corresponding roughly both with historical development, as well as the amount of ``intelligence'' added by learned components.
\begin{enumerate}
\item
\textbf{Primary utilization in DBMS:}
Many available commercial and open-source DBMS have a long tradition of using collected statistics and statistical estimators to support some of their data management functions.
Some of these methods serve as an integral parts of some components. For example, cardinality estimation (\CE) consists of estimating data distribution properties used later as input for query optimization (QO). While for instance SQL Server and \postgres build histograms and collect frequently occurring values for table attributes, MySQL and MariaDB apply a sampling strategy for \CE. With respect to non-integral functions, basic statistical methods are used for advisory functions, such as index or view advisors, or knob tuning and SQL rewriters~\cite{zhou2020database}.

\item
\textbf{Individually learned components:}
With the recent increased interest in machine learning, an increased amount of work has tried to design ML-based models to replace certain components in DBMS. In this context, the most representative works address QO and indexing. For QO, a variety of supervised and unsupervised models have be proposed for \CE~\cite{han2021CEbenchmark, wu2020bayescard, hilprecht2019deepdb, yang2019deep, zhu2020flat, hasan2019multi}, cost estimation~\cite{siddiqui2020cost, vu2021learned} and join order selection~\cite{marcus2018deep, sabek2021case}. Meanwhile, a number of learned data structures structures are proposed for single and multi-dimensional indexing~\cite{kraska2018case, nathan2020learning, perera2021no}. It has been shown such ML-based methods often exhibit superior performance, when evaluated in numerical experiments outside of actual systems or tightly controlled experimental setups inside of real systems (e.g. through injecting cardinality from an outside file). However, many problems relevant to real-world deployment, such as e.g. explainability vs. predictive power, robustness or fault tolerance remain largely unaddressed.

\item
\textbf{Comprehensively learned modules:}
On top of individually learned components, current work further moves forward by trying to substitute an entire functional module, e.g. the QO, executor or storage engine with a machine learning model. Some approaches combine multiple learned components together, learn to steer existing modules~\cite{marcus2020bao}, or even
attempt to "learn" an entire module in an end-to-end fashion~\cite{marcus2019neo}. Although most work of this type claims to achieve incredible performance gains, many proposed solutions appear to be impractical for actual deployment in real-world DBMS. The reasons for this are two fold. 
First, learned modules of this type are highly task specific and data dependent, which may cause serious shortcomings, including but not limited to: cold start problems, lack of generalization, tuning difficulties. 
Second, replacing an entire module may cause compatibility risks and may require substantial engineering efforts. Therefore, at this stage, even evaluating a learned module inside a real DBMS is a difficult task.

\item
\textbf{AI-Native databases:}
Some very recent work even proposes to redesign the whole architecture of DBMS to fully adapt AI models for data management. For example,~\cite{wu2021unified} proposes a "one model for all architecture", which learns a shared representation of data and query knowledge and then fine-tunes smaller models for each specific task.~\cite{li2019xuanyuan} proposes  "AI-designed databases", where AI  is integrated into the life cycle
of database design, development, evaluation, and maintenance, hoping to provide optimal performance for every scenario. Although this appears to be an intriguing vision, it is clear they are still far away from any useful implementation.

\end{enumerate}

Hence, regarding ``AI for DB'', we observe that:

1. This field is very prosperous. The research efforts range from individual components to complete modules to even the whole architectures, and include many scenarios (QO, indexing, execution, storage and etc). Enough evidence has shown that AI-based solutions could indeed improve  database performance and have the  potential to play significant roles in next-generation DBMS. 

2. Current work mainly focuses on ``AI'' solutions but is not concerned with how to actually deploy them ``for DB''. In addition to that, more realistic tests for newly proposed methods relying on e.g. existing extension functionalities of DBMS are rarely described in the literature, possibly because it requires deep expertise beyond the design of ML models. Such expertise requires a comprehensive understanding of both AI and DB perspectives and a systematic co-design of both algorithm and systems.

Therefore, we believe that a SysML framework such as \baihe, which supports both evaluation and deployment of AI-driven solutions in real systems is crucial for the further development of the AI for DB field.


\subsection{High Level Architecture of \baihe: Fundamental Design Choices and Trade-Offs}
\label{subsect:high_level}
\begin{figure*}[h]
  \centering
  \includegraphics[width=0.65\linewidth]{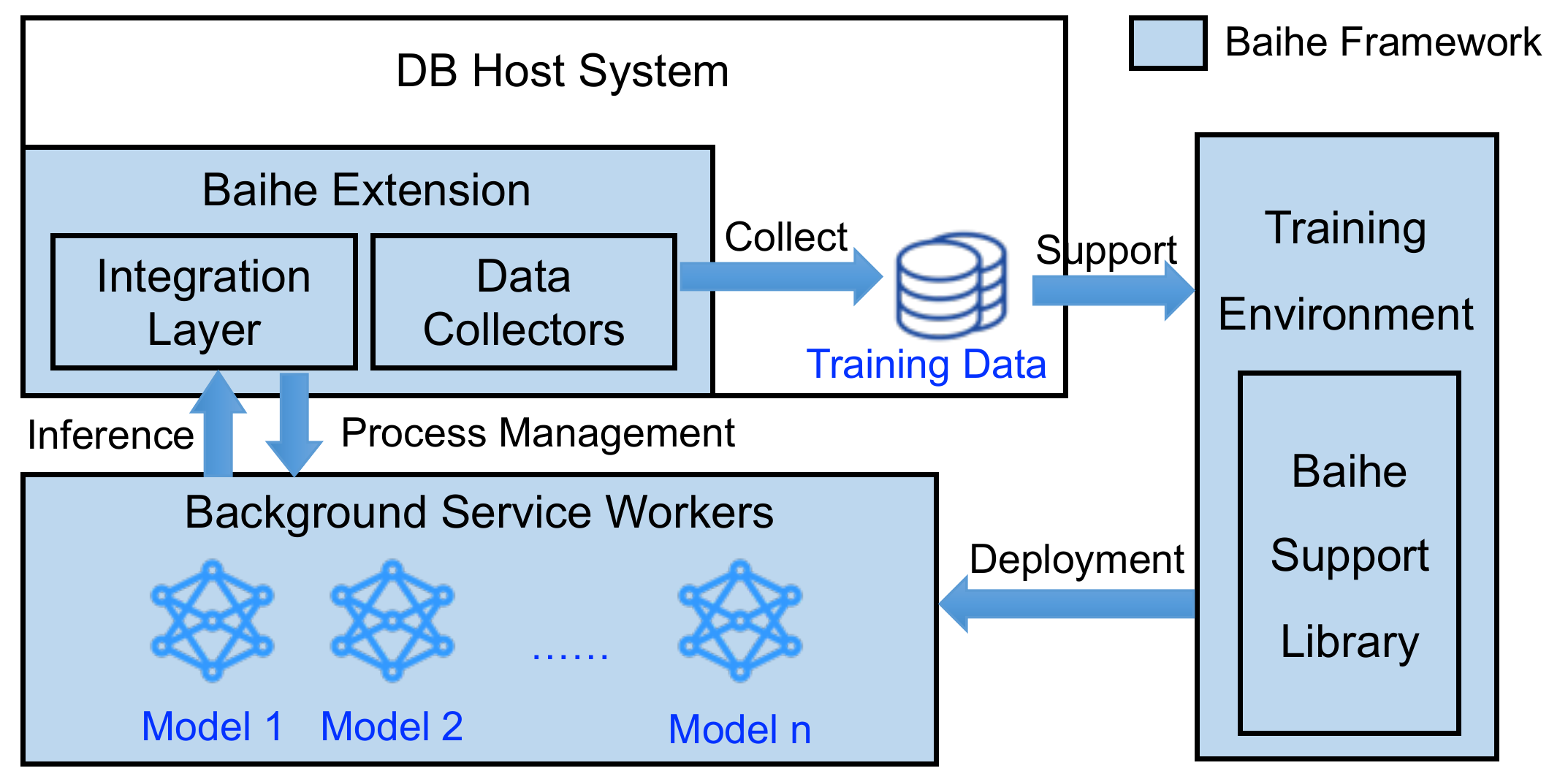}
  \caption{High Level View of \baihe's Architecture.}
  \label{fig:arch}
 \end{figure*}
We now briefly describe the high level design of \baihe and discuss the rationale and design philosophy behind it. We note that this design should rather be considered as a design blueprint resp. design pattern, which may then have to be adapted to a specific host database system. As an example we present \baihe for \postgres in Section \ref{sect:design}.

The core part of \baihe is an extension which plugs directly into a database host system through an existing extension mechanism. This extension is the central component of \baihe, as it
\begin{enumerate}
    \item intercepts the query planning and execution process, so that individual steps may be substituted with model inference calls, 
    \item contains a clone of the host system's planner ("shadow planner"), which can be conveniently extended and modified, so that core functionality may be overwritten without interfering with the host system itself,
    \item provides convenient configuration facilities,
    \item controls collection of training data.
\end{enumerate}
Most of the above functionality is encapsulated in the Integration Layer component, which is the central control unit for \baihe. 

Model inference is decoupled from the extension itself, this means that for every model a new background worker process is used. These processes communicate with the \baihe extension and return results for inference requests. If the host system supports it, the control of these processes is managed using the host systems process management capabilities. 

Finally, \baihe needs a support library. This component allows for simple access to collected training data, such as e.g. query runtime statistics or saved query plans. Furthermore, it provides functionality for convenient deployment of trained models into the host system.

The design above is based on the following high-level requirements:
\begin{itemize}
    \item Separation from the core system.
    \item Minimal third party dependencies.
    \item Robustness, stability and fault tolerance.
    \item Usability and configurability.
\end{itemize}
For the remainder of the section we now describe the impact of these requirements on our design in more details. 
\newline

\textbf{Separation from the Core System.} 
Both commercial and open source RDBMS have been and are being actively maintained and supported over long periods of times. Commercial license and support agreements may span years or even decades, so that customers may receive regular updates and critical patches, especially those addressing security issues. In order to impact existing processes as little as possible, all while avoiding a development of an entirely new database system, we have chosen to develop \baihe as an extension on top of an existing system, so that we may keep it as separate from the host system as possible. In this way, we avoid unnecessary doubling of maintenance efforts and allow for a quicker pace of development.

\textbf{Minimal Third Party Dependencies.} 
Modern ML stacks are characterized by a large number of dependencies, comprising low level numerical libraries and GPU kernels, intermediate ML framework and runtime codes (typically implemented in C++), as well as high level integration code typically written in Python. To train and use what may now even be considered relatively simple deep models, a large number of packages and libraries at these three levels must be present on a host system. In practice, this typically leads to a high degree of maintenance efforts for both production, as well as test and development environments. This is particularly problematic in more  traditional organizations, such as financial institutions, where - due to security or legal reasons - individual software packages  may have to go through a specific vetting and approval process. We have thus designed \baihe such that it requires a minimal amount of external dependencies beyond those required by the core system. 

\textbf{Robustness, Stability and Fault Tolerance.} For most modern applications databases are among the most fundamental and mission-critical components. It is thus imperative that additional deployment of ML-based components into such systems does not impact existing service level agreements or interfere with related operational requirements. Therefore, errors arising from e.g. model-based predictions, which might influence an individual session or the system as a whole should be detected and mitigated through fallbacks to existing core functionality.

\textbf{Usability and Configurability.} To further ease the burden of integration, it should be possible to control and configure \baihe through standard mechanisms offered by the host system. In the concrete case of a system such as  \postgres this means that \baihe should be able to be configured through the usual \postgres configuration files, as well as provide a set of user defined functions as well as stored procedures, such that \baihe functionality may be configured, activated or deactivated through any authorized  command session and without requiring any restarts of the system as a whole.

Based on the above four points we may now formulate more concrete design goals with respect to the machine learning aspects.

\textbf{Model Support.} As a general framework, \baihe should support deploying models for a range of different tasks, such as e.g. cardinality estimation, join order selection, query run time prediction. Furthermore, it should support models for learned data structures, such as e.g. learned indices.

With respect to the models themselves, \baihe should offer support for both neural network, resp. deep learning based, model families, as well as more traditional ones, such as e.g. probabilistic graphical models, decision trees, random forests or gradient boosted trees.

\textbf{Model Training.} 
As we have discussed in the previous section - despite the recent progress in AutoML~\cite{zhou2020database} - model training still requires close expert supervision supported by suitable tooling allowing for thorough model evaluation and rapid experimentation. In \baihe we thus prefer to decouple model training from the rest of the system as much as possible. While \baihe should still provide suitable functionality for training data collection, as well as tools for model export and deployment, we believe that training itself should be set up in a separate environment under control by specialist users such as data scientists or machine learning engineers. Once training has achieved satisfactory progress, a model can then be deployed using a well-defined deployment process.

\textbf{Model Inference and Deployment} 
In order to avoid expensive serialization and de-serialization steps one might want to integrate a model directly into e.g. the planner component of the host system. On the other hand, to maintain the maximal amount of flexibility with respect to software dependencies and computing resources needed for inference, one might also consider implementing model inference in a completely separate service process outside of the control of the host system. We believe that both of these extremes would clash with requirements on robustness and stability, as well as maintenability of the host system. In \baihe we thus choose to isolate inference in a separate process, but keep this process under management by the host system. To eliminate the need for expensive serialization steps we furthermore propose process co-location, so that existing shared memory facilities may be used as much as possible. 

While current (practical) models in the SysML space, even deep ones, can still be considered relatively light-weight~\cite{han2021CEbenchmark, zhu2020flat}, we hence believe that computational resources needed by the inference process would in general not adversely affect the core databases process on the same machine. In the long term, should there be the need for computationally more expensive models to be deployed in the system, one could address this problem, at least in cloud environments, through on-demand attachable resources.

As a consequence of the co-location requirement, one could imagine having to install additional packages on the machine running the host system, which would be needed to run model inference (e.g. ML framework runtimes etc.). To address this issue \baihe should provide proper tools allowing for exporting models such that they can readily be used with as little extra dependencies as possible. To achieve this goal, we propose to make use of the recent advances in the context of ML model compilers~\cite{ben2019modular}, which make it possible to compile models together with custom CPU or GPU based math kernels into highly efficient binary code, that may be accessed through a C-ABI. Nevertheless, \baihe should support a "Development Mode", where models may be developed and tested in the host system without intermediate compilation and build steps. 

\section{System Design and Implementation}
\label{sect:design}
We have started \baihe development as a first step towards the long term vision outlined in Section \ref{sect:learned_db}. As such, our main target for \baihe has been deployment in the widely used \postgres Database System, which - thanks to its open nature and high degree of extensibility - makes for an excellent candidate system for retrofitting modern ML-based approaches into a real-world DB. We now describe \baihe's design and implementation for \postgres.

\baihe for \postgres consists of three components: The \baihe extension for \postgres, a number of \baihe background worker processes and the \baihe support library. The relation between these components is shown in Figure \ref{fig:arch}. We now describe these components in more details.

\textbf{\baihe extension for \postgres.} This component is loaded upon database startup and uses the existing hooking mechanism of \postgres in order to intercept query planning and execution. All of \baihe's functionality is controlled through this component, including collection of training data (e.g. query runtime statistics) and model handling. The entry point for all of these functions is implemented in the \textbf{Integration Layer} component. All outside calls first end up in this component. Upon startup, it will first request additional shared memory from the \postgres host system where it initializes control data structures needed for both the \textbf{IPC Module} and \textbf{Data Collection} module. 

\textbf{Background Workers.} Once a model is available for deployment, \baihe can be requested to begin using it. In this case it will use the existing \postgres functionality to start a background worker process which will load the deployed model from the file system and wait for incoming inference requests on a message queue living in a shared memory space. 

\textbf{\baihe Support Library.} The \baihe support library provides a high level Python API which serves two main purposes: Provide access to training data collected by the \baihe extension and furthermore expose functionality needed to deploy models in either production or development mode. Model deployment in development mode amounts to saving model weights and creating a Python module which can loaded by a Python interpreter embedded in a \baihe background worker process. In production mode a custom LLVM-based model compiler is used to compile the entire model into a standard shared library, that is loaded dynamically by background workers.

\begin{figure*}[h]
  \centering
  \includegraphics[width=0.8\linewidth]{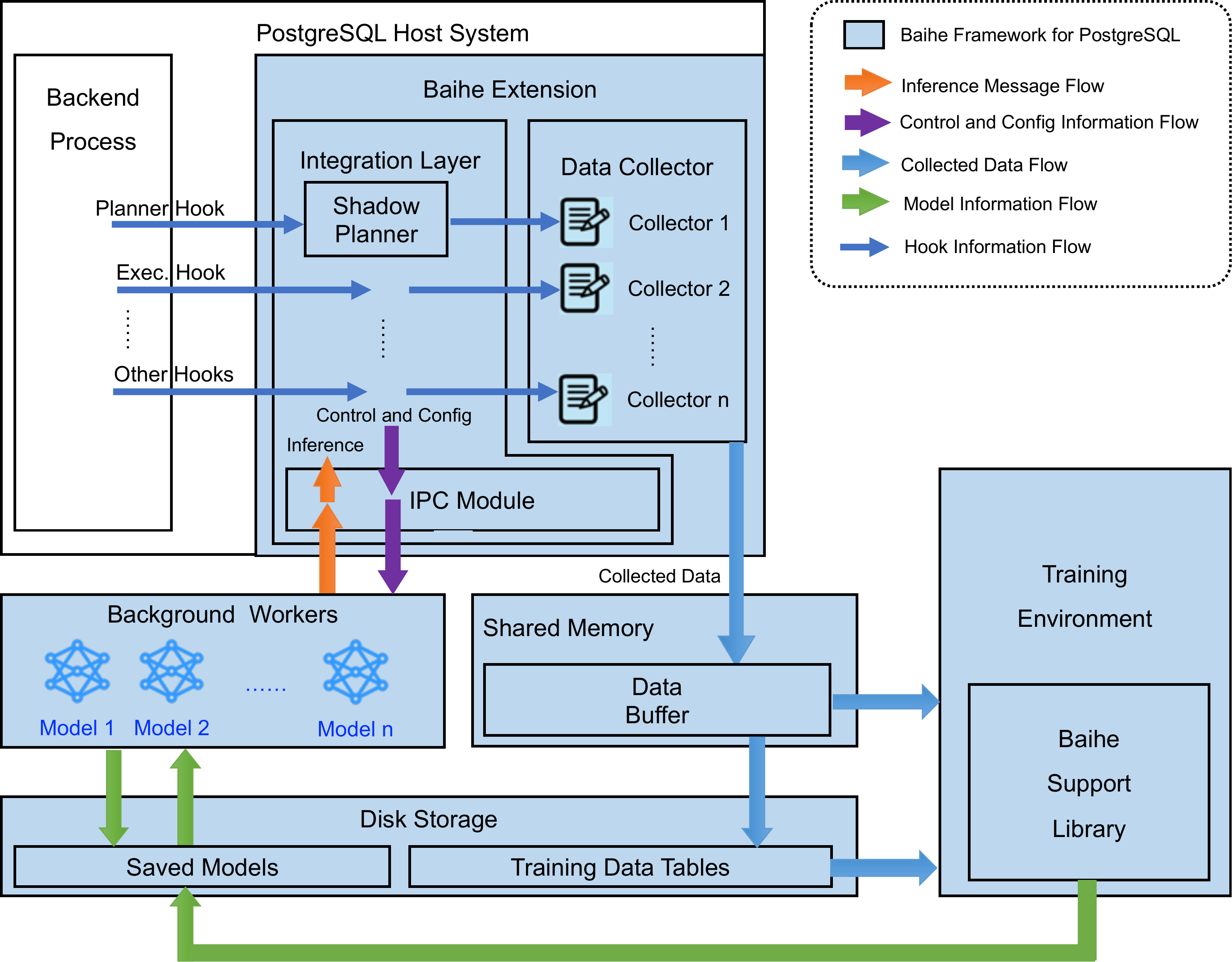}
  \caption{\baihe Architecture.}
  \label{fig:arch}
 \end{figure*}

\subsection{\baihe Integration Layer and IPC Module}

The \baihe integration layer is the central control unit of the \baihe extension: It is accessed from the host system by implementing some of the hooks already defined in \postgres, where it implements \baihe's high level logic. All functionality for communication with background workers, as well as necessary process management for background workers is encapsulated in the IPC Module. The IPC module makes extensive use of Postgres core APIs used for management of shared memory, as well as process management.

Additionally, the \baihe Integration Layer defines user defined functions and session variables which are necessary for controlling model handling and data collection, e.g. starting and stopping background workers, defining which models should be used in which situations, as well as defining when and how training data should be collected.

\textbf{Shadow Planner.} 
To allow for a maximum degree of flexibility with respect to models providing input for  query planning, the \baihe extension contains a customized duplicate of the core \postgres planner as a  "Shadow Planner" component. In this way we can achieve the following: First, we may freely add new hooks into the planner without having to modify core source code\footnote{Currently, \postgres itself offers e.g. no hooks to overwrite cardinality estimation for e.g. single table queries or join size estimates}. Second, the behavior of the planner as a whole may be adjusted and new ideas tested without having to interfere with the core source. As an additional benefit, this reduces overall compilation and build times. 

Currently, in addition to the existing hooks originating from the original \postgres code, we have equipped the shadow planner with the following hooks:
\begin{itemize}
    \item Cost Model: we add an additional hook per "node" in a query plan. This allows for overwriting cost predictions for such operations as sequential scan, index scan, nested loop join, hash join etc. Furthermore, we incorporate hooks for estimating costs of query predicates making use of operations beyond comparison operators for numerical values. We plan to further support hooks for overwriting the cost estimation of user defined functions etc.
    \item Cardinality Estimation: We add hooks at several levels of the cardinality estimation process, such as cardinality estimation for a query involving a single table or a join between two tables.
\end{itemize}

To improve the interplay between hooks and planner code, we furthermore design the concrete hook signatures and calling code with error handling mechanisms allowing for seamless fallback to standard planner behavior in the case of errors originating e.g. from erroneous model inference calls. We discuss how shadow planner and related hooks are specifically used for query optimization task in Section~\ref{sect:example}.

\subsection{Data Collection}
The design of \baihe's data collection module borrows heavily from the popular \textsf{pg\_stat\_statements} extension for \postgres. However, since \baihe targets data collection for training machine learning models, it takes a more dataset-centric point of view. 

More concretely, \baihe data collection is designed around the notion of "Data Collectors". Users may define and activate several Data collectors at the same time, where each data collector may be defined as a set of filter conditions plus a versioned data set identifier. In this way users may for each Data Collector specify the following:
\begin{itemize}
    \item \textit{Filter conditions}: For which query type (SELECT, INSERT, ...) involving which tables should this data collector be applied?
    \item \textit{Dataset identifier and version}: For reproducibility a debugging purposes it is essential to keep track of exactly which data was used to train a specific model version. Hence, any data set collected is identified through both a dataset identifier as well as a version number.
    \item \textit{Features}: For some dataset and model combinations, only queries themselves might need to be collected, while for others it might be necessary to also collected generated query plans, together with estimated costs and actual run times both on query plan, as well as plan-node level. Users may flexibly specify, which features should be saved by a data collector
\end{itemize}
Data collection may be controlled entirely through a standard command session. After a data collector has been defined through a call to a \baihe stored procedure, the data collection process itself may also be started and stopped by running start and stop routines exposed from the \baihe extension through custom stored procedures. See Figure \ref{fig:usage_data} for an example.

While a Data Collector is active, all collected data will be stored in shared memory.In the case of very large datasets, shared memory content may be temporarily stored on disk. Once data collection is stopped, a data set with incremented version identifier will be written to disk and made available in a table specified in the Data Collector's configuration. Training data may then easily be accessed through SQL.

\subsection{Model Integration}
\begin{figure*}[h]
  \centering
  \begin{subfigure}{8.5cm}
  	{\color{darkgray} \textsf{\#\# Define a data collector}} \\
  	{\color{darkgray} \textsf{\#\# Filter queries by tables and query type}} \\
    \textsf{CALL {\color{blue} DEFINE\_DATA\_COLLECTOR} ( ``CardEstCollector'', \\
    \{ ``tbl\_users'', ``tbl\_items'', … \}, \{ ``SELECT'' \} ); } \\

  {\color{darkgray} \textsf{\#\# Start data collection}} \\
  \textsf{CALL {\color{blue}START\_DATA\_COLLECTOR} ( ``CardEstCollector'', \\
  	``Data\_Set\_1'', ``tbl\_training\_data''  ); } \\
  
  {\color{darkgray} \textsf{\#\# Stop data collection}} \\
  \textsf{CALL {\color{blue}STOP\_DATA\_COLLECTOR} ( ``Data\_Set\_1''); } \\
  
  \caption{Configuring data collection for a single cardinality estimation model. Only data related to SELECT queries touching certain tables is collected. Data collection can be started and stopped.}
  \label{fig:usage_model}
  \end{subfigure}
  \hfill
  \begin{subfigure}{8.5cm}
  	{\color{darkgray} \textsf{\#\# Model Registration}} \\
  	\textsf{CALL {\color{blue} REGISTER\_MODEL} ( ``MyCardEstModel'', ``CARDEST'', \\
  		\{ ``tbl\_users'', ``tbl\_items'', … \}, ``tbl\_my\_cardest\_model\_stats'' ); } \\
  	
  	{\color{darkgray} \textsf{\#\# Start Background Process for Model}} \\
  	\textsf{CALL {\color{blue} START\_MODEL} ( ``MyCardEstModel'' ); } \\
  	
  	{\color{darkgray} \textsf{\#\# Stop Background Process for Model}} \\
  	\textsf{CALL {\color{blue} RESET\_MODEL} ( ``MyCardEstModel'' ); } \\
  	
  \caption{Deploying a trained model into the system: The model is used only for queries touching certain tables and maybe activated or deactivated when requested by the user.}
  \label{fig:usage_data}
  \end{subfigure}
  \hfill
  \caption{Example configuration sessions for \baihe}
  \label{fig:usage}
\end{figure*}
Along the lines of our requirements on minimization of dependencies and model inference and deployment as described in Section \ref{sect:background}, model inference takes place in background worker processes. For every model registered in \baihe, a user can request \baihe to start a background worker process, which will 
\begin{enumerate}
    \item Load the saved model from disk
    \item Connect to the \baihe shared memory space
    \item Wait for incoming inference requests on a message queue.
    \item Once an inference request is received, the background worker will run the request through the loaded model and return inference results (which may possible also just a flag indicating that an error has occurred).
\end{enumerate}

Once a query is submitted by a client to the corresponding \postgres backend process, the query planning and execution process will be intercepted by the \baihe extension and depending on model type, a number of inference requests will be sent to the correct background workers. All communication is implemented asynchronously, so that a backend may fall back to standard functionality in case a background worker is not available. 

Out of the Box \baihe allows for the integration of custom models for query runtime prediction, as well as cost and cardinality estimation. Models of these types may be used directly without changes to the \baihe extension source code. More specific types of models, requiring e.g. new hooks at certain places in the planner code,  may easily be supported with slight changes to the \baihe extension code.

Similar to the data collection functionality described in the previous subsection, \baihe's model handling facilities are controlled and configured using a number of stored procedures and user defined functions implemented in the \baihe extension. A simple usage example is displayed in Figure \ref{fig:usage_model}: Through a standard command session users with the right permissions may request models to be registered and the corresponding background workers to be started or stopped. This allows for model updates without having to restart the entire system. 

As mentioned previously in Section \ref{sect:background} \baihe's focus is on model inference only. This means that the training process itself, that is solving the optimization process for a certain combination of model and training data, does not run in any \baihe components. Instead, the usual development process can be outlined as follows: \begin{enumerate}
    \item Training data is selected and downloaded using the \baihe support library, implemented as a Python packages. 
    \item A model can then be defined and trained, preferably using a framework supported by \baihe's production mode. Currently supported frameworks are sklearn and Tensorflow. Training is controlled entirely by an expert user, such as e.g. a data scientist or machine learning engineer.
    \item Once the model has been trained and evaluated, it may be deployed using the \baihe support library.
    \begin{itemize}
        \item \textit{Development mode}: In this mode, the model is deployed as a Python model on the database servers file system, together with an automatically created environment containing all the model's dependencies. A background worker then uses an embedded Python interpreter to access the model. 
        \item \textit{Production mode}: In this mode, \baihe's support library is used to compile the model including trained parameters into a shared library that is loaded dynamically by a background worker. The shared library does not depend on any external numerical or ML framework libraries. Model code itself will be compiled together with a number of math kernels (implementing e.g. matrix multiplications, convolutions, etc.) into a self contained component with a standardized interface. 
    \end{itemize}
\end{enumerate}

\section{Example Use Case: Learned Query Optimizer for \postgres}
\label{sect:example}

We describe now a typical use case of for \baihe: deploying a learned query optimizer into \postgres. More concretely, we discuss here the following two variants:
\begin{enumerate}
    \item QO with individual components: cardinality estimation and cost model substituted with separately trained components.
    \item End-2-End QO: Here, the entire query optimizer is substituted by a trained model.
\end{enumerate}

\subsection{QO with individually learned components}
\CE models and ML-based cost models are supported out-of-the box. The shadow   planner in the \baihe integration layer intercepts all requests for a cardinality estimate for a query touching a set of tables $T$ with set of query predicates $Q$. The tuple $(T,Q)$ is obtained from internal query parse tree structures, serialized and passed into a trained model running in a background worker. The model then returns a selectivity $0 \leq s \leq 1$, which is passed on to the planner.

Training data collection for a \CE model depends on whether the model is based on query-driven or data-driven \CE. Specifically, data-driven \CE methods build unsupervised models over the tabular data, then the cardinality of any query could be estimated over this model. For data driven \CE no additional data collector is needed, since models may be trained simply using (samples of) table data and schema information provided by the user (the latter being required for models supporting multi-table \CE).

Query-driven \CE methods build the supervised models mapping featurized queries to the cardinality. For a query driven model, we first define a data collector, which, for every query, saves the entire query plan, together with all statistics collected during the execution (i.e. the entire output of  EXPLAIN ANALYZE). Training code can then load this data and convert it to the required form of $(\text{Subquery}, \text{Cardinality})$ records needed for training.

Once a \CE model has been trained it may then easily be registered as a model of "CARDEST" type using a call to the corresponding \baihe procedure. To make the model active, a background worker is started using another \baihe procedure call and the session variable "baihe\_ce\_model" is set to the model identifier. Then, all subsequent queries for this session will use learned cardinality from the deployed model. 

The process works similar in the case of cost models. Out of the box, the \baihe will shadow planner intercepts all cost-estimation call on a node-level (e.g. sequential scan, index scan, etc.). Then, a record depending on a variable number of features (depending on node type), is built and sent to a cost estimation model running in a background worker, which then returns a predicted cost in terms of cost units.

To collect  training data, we register a data collector with the same settings as used for query-driven \CE. In this way, for every node in a query plan we obtain all features required for training a meaningful model, with the most important features being node type, estimated cardinality, actual cardinality and the time needed to execute a node.

\subsection{End-2-End Learned QO}

Some recent work also presents methods for learning a query plan directly. Such methods take a query as input, apply a certain featurization scheme and return an entire query plan as output. For our example we take a closer look at two major representatives of this line of work, namely NEO~\cite{marcus2019neo} and BAO~\cite{marcus2020bao} and  show how they could be deployed using \baihe.

NEO applies tree convolution networks to extract features from structured query plans and learns a function, called value network, mapping plans to execution latency. Then, a best-first search strategy is used to find a near-optimal query plan as measured by the value network.

To deploy NEO, we first register a data collector as used for query-driven \CE in the previous subsection. Then, we a background workers which implements value network inference and the best-first search strategy, respectively. During query execution, we intercept the planning process at the highest level in the \baihe integration layer (right after the \baihe extension is first called by the host system) and forward the query to the the background worker, evaluates the value network worker and returns a query plan after running the best-first search. This plan is directly sent to the \postgres engine for execution.

BAO adapts a different strategy than NEO. It learns to steer but not the replace the QO. Specifically, it also learns a latency prediction network which maps a query plan to its execution latency. For each query, it generates several plans corresponding to different hint sets and then selects the plan with the minimum predicted latency. Hence, to deploy BAO, data collection and model deployment need to be configured in exactly the same way as BAO and NEO.

Note that the above discussion only concerns the case where both the BAO and NEO models have been trained to a certain point and then remained unchanged after deployment. However, both models have been designed to be updated in an online-manner, so that they may possibly adjust to changes in the underlying data and workload, without having to be explicitly retrained. 

While online updates are not directly supported yet, we note that model code running inside background workers could easily be written in such a way that incoming inference request data may simultaneously be used to updated the model running inside a worker. However, we note that - at least for now - it is then the responsibility of each such background worker to properly manage  model state, ensure that model updates don't block future inference requests and deal with errors that might occur during online updates.

\section{Open Source Release and Future Plans}
\label{sect:conclusion}
As development of \baihe has started only recently, it is not yet available for general use. However, we plan to release a first version of \baihe for \postgres under an open source license in the beginning of 2022. This version should contain all of the essential functionality needed to build and experiment with learned query optimizers as described in the previous subsection. Later on, in the second half of 2022, we plan to release an extended version of \baihe which has seen first tests under real world conditions and supports production-mode deployments. 

The reasons for this release schedule are as follows:
\begin{itemize}
    \item First, We hope to encourage community participation in the development of \baihe as soon as possible.
    \item Second, We wish to serve the DB research community by providing a flexible and easy to use experimental platform for for future research into AI4DB, hoping to establish a standardized and realistic test bed for future models and algorithms.
\end{itemize}

Overall, we hope to have provided convincing arguments for the soundness and practicality of \baihe as a design blueprint. The ongoing development of \baihe for \postgres should further help refining this blueprint and serve as an implementation guide for other database systems.

Besides the ongoing development, there are many avenues for future work. For instance, the current version of \baihe has been designed with most applications revolving around query optimization. However, one could envision \baihe to be used in the context of learned indices, database configuration tuning or other advisory functions.

Another aspect that has been left out for now concerns the training process itself, as well as models which may benefit from online training. Integrating  training and online updates of possibly large models directly into \baihe should certainly provide for many interesting system design challenges.

Finally we note that the development of production mode deployment needs a custom model compiler infrastructure, which further adds to the many engineering and research challenges that accompany this line of work. We encourage the entire community to actively participate and accept some of these challenges. 

\bibliographystyle{ACM-Reference-Format}
\bibliography{main}
	
\end{document}